\documentclass[prb,showpacs]{revtex4}
\usepackage{graphicx,psfrag,amssymb,feynmf}

\begin{document}
\bibliographystyle{apsrev}

\title{Threshold electric field in unconventional density waves}
\author{Bal\'azs D\'ora}
\affiliation{Department of Physics, Technical University of Budapest, 
H-1521 Budapest, Hungary}
\author{Attila Virosztek}
\affiliation{Department of Physics, Technical University of Budapest, 
H-1521 Budapest, Hungary}
\affiliation{Research Institute for Solid State Physics and Optics, P.O.Box 49,
H-1525 Budapest, Hungary}
\author{Kazumi Maki}
\affiliation{Department of Physics and Astronomy, University of Southern 
California, Los Angeles CA 90089-0484, USA}

\date{\today}

\begin{abstract}
As it is well known most of charge density wave (CDW) and spin density wave
(SDW) exhibit the nonlinear transport with well defined threshold electric
field $E_T$. Here we study theoretically the threshold electric field of
unconventional density waves.
We find that the threshold field increases monotonically with temperature
without divergent behaviour at $T_c$, unlike the one in conventional CDW.
The present result in the 3D weak pinning limit appears to describe rather well
the threshold electric field observed recently in the low-temperature phase
(LTP) of $\alpha-(BEDT-TTF)_2KHg(SCN)_4$.
\end{abstract}


\maketitle

\section{Introduction}

A striking feature of superconductors discovered after 1979 is that they
are mostly unconventional \cite{szupravezetes}. The case of d-wave superconductor for both the
hole-doped and electron doped high $T_c$ cuprates is now well
established \cite{bla1,bla2,bla3,bla4,bla5,bla6}. Also most of heavy fermion superconductors and organic
superconductors appear to be unconventional \cite{bla7,bla8,bla9,bla10}.

Therefore it is very natural to consider unconventional density waves (UDW)
within this general context. Recently a model of unconventional SDW was proposed and its
thermodynamics and optical properties were studied \cite{kiscikk,nagycikk}.

The object of this work is to study the threshold electric field of UCDW
and USDW associated with the Fr\"ohlich conduction of UDW. This is motivated by
the threshold electric field $E_T$ measured in the low-temperature phase (LTP) of
$\alpha-(BEDT-TTF)_2KHg(SCN)_4$ \cite{ltp}, where the LTP appears not to be conventional
DW. There is no X-ray or NMR signature characteristic to conventional CDW or SDW. 
$E_T$ in this salt increases monotonically
with increasing temperature somewhat similar to the one observed in SDW of
Bechgaard salts $(TMTSF)_2PF_6$ \cite{tmtsf1,tmtsf2,tmtsf3}. However the details are quite
different. At low temperature the observed $E_T$ increases linearly with
$T$. Also the enhancement at $T_c$ is much larger than the one observed in
SDW of Bechgaard salts. The nature of the LTP of $\alpha-(ET)_2KHg(SCN)_4$
is not well understood in spite of many studies on the magnetoresistance,
Schubnikov-de Haas effect and the Haas van Alphen effect \cite{osszefoglalo}. 
Roughly speaking
 $\alpha-(ET)_2$ salts 
may be put into two groups: one superconducting and another with this
mysterious LTP.

It appears that $\alpha-(ET)_2MHg(SCN)_4$ with $M=K$, $Tl$ and $Rb$ belong
to the group with the LTP. At least the sensitivity of the LTP to magnetic
field indicates that the LTP is not a SDW but a kind of CDW \cite{et2_1, et2_2}.
 Indeed the
$H-T$ phase diagram of the LTP in $\alpha-(ET)_2KHg(SCN)_4$ determined by
magnetoresistance measurement is very similar to the one of
Fulde-Ferrell-Larkin-Ovchinikov (FFLO) state \cite{fflo1,fflo2} 
in a d-wave superconductor \cite{fflomaki}. The
FFLO in a d-wave superconductor extends to much higher magnetic field than
the one in s-wave superconductors \cite{fflo1,fflo2}.

If we assume that the Pauli paramagnetism is driving the magnetic phase
transition, the $H-T$ phase diagram of UCDW is the same as the one in
 a
d-wave superconductor.
 Also we shall see later that $E_T$ in UCDW describes 
well the threshold
electric field observed in the LTP of $\alpha-(ET)_2KHg(SCN)_4$. Therefore
we may conclude that the LTP of some of $\alpha-(ET)_2$ salts is UCDW.

\section{Phase Hamiltonian and the threshold electric field}

In terms of the phase $\Phi({\bf r},t)$ of DW the phase Hamiltonian is given by 
\cite{tmtsf1,tmtsf2}
\begin{eqnarray}
H(\Phi)={\bf \int}d^3r\left\{\frac 1 4 N_0 f \left[ v_F^2 \left(\frac{\partial\Phi} {\partial x}\right)^2
+v_b^2 \left(\frac{\partial\Phi} {\partial y}\right)^2
+v_c^2 \left(\frac{\partial\Phi} {\partial
z}\right)^2+
\left(\frac{\partial\Phi}
{\partial t}\right)^2-4v_FeE\Phi\right]+V_{imp}(\Phi)\right\},
\end{eqnarray}
where $N_0$ is
the density of states in the normal state at the Fermi surface per spin,
$f=\rho_s(T)/\rho_s(0)$ where $\rho_s(T)$ is the condensate density and $E$
is an electric field applied in the $x$ direction. Here $v_F$, $v_b$ and
$v_c$ are the characteristic velocities of the quasi-one dimensional
electron system in the three spatial directions. For UDW the condensate
density is the same as the superfluid density in d-wave superconductors \cite{d-wave}.

Now let us consider $V_{imp}(\Phi)$, the pinning potential due to impurities. It is
immediately clear that if we consider point like scatterers (s-wave), the
potential would be zero at every order in the impurity scattering due to the zero average of the
gap. Beyond this approximation, one can
take other wave vector dependent terms into account, originated from an
expansion in terms of Fermi surface harmonics, which are plane waves in our
quasi-one dimensional system. Indeed such a model has been
introduced by Haran and Nagi \cite{haran} in order to describe the defects introduced in
high $T_c$ cuprates by electron irradiation. In fact this model is
successfully applied to formulate the upper critical field of the
electron-irradiated $YBCO$ \cite{ybco1,ybco2,ybco3}. 

  The form of the important
matrix element (with wavevector close to the nesting vector) reads as
\begin{equation}
U({\bf Q+q})=V_0+\sum_{i=y,z}V_i\cos(q_i \delta_i),
\end{equation} 
where the higher harmonics are neglected because of their smaller
coefficient. The first order term in the pinning potential vanishes because of the wavevector
dependence of the gap in UCDW while in USDW it vanishes already due to the
sum over spins. In the followings we assume that the gap of UCDW is given by
$\Delta({\bf k})=\Delta\cos(k_y b)$. Note that we can obtain identical
results with $\Delta({\bf k})=\Delta\sin(k_y b)$ and for a gap dependent on
$k_z$ as well.

\vspace*{1cm}
\unitlength=1mm

\begin{figure}[h]
\hspace*{7cm}
\scalebox{0.7}{
\begin{fmffile}{etmf}
\begin{fmfgraph*}(40,25)
{
 \fmfleft{v1}\fmflabel{$U(\bf Q-k+k^\prime)$}{v1}
 \fmfright{v2}\fmflabel{$U(\bf Q+k-k^\prime)$}{v2}
 \fmf{fermion,left,tension=0.3,label=$i\omega_n$}{v1,v2}
 \fmf{fermion,left,tension=0.3,label=$i\omega_n$}{v2,v1}
 \fmfiv{l=$\bf k^\prime$,l.a=135,l.d=.55w}{c}
 \fmfiv{l=$\bf k^\prime-Q$,l.a=45,l.d=.55w}{c}
 \fmfiv{l=$\bf k$,l.a=-45,l.d=.55w}{c}
 \fmfiv{l=$\bf k-Q$,l.a=-135,l.d=.55w}{c}
 \fmfvn{decor.shape=cross,decor.size=4thick}{v}{2}
}
\end{fmfgraph*}
\end{fmffile}}
\vspace*{1cm}
\caption{The diagram of the lowest order contribution of impurities to the pinning potential
 is shown. The solid line denotes the electrons, the crosses denote the impurities. \label{fig:lowest}
 }
\end{figure}
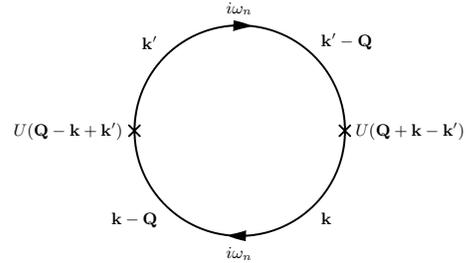

 The lowest order nonvanishing diagram contains a closed loop with two crosses 
of impurities (see Fig. \ref{fig:lowest}), and the pinning potential is obtained as
\begin{eqnarray}
V_{imp}(\Phi)=- \frac{8 V_0 V_y N_0^2}{\pi}\sum_j \cos(2({\bf
QR}_j+\Phi({\bf R}_j))) 
\Delta(T)\int_0^1\tanh\frac{\beta\Delta(T)
x}{2}E(\sqrt{1-x^2})(K(x)-E(x))dx,
 \label{pinningpot}
\end{eqnarray}
where $\Delta(T)$ is the temperature dependent order parameter 
\cite{nagycikk,d-wave}, ${\bf R}_j$
 is an impurity site, $K(z)$ and $E(z)$ are the complete elliptic
 integrals of the first and second kind, respectively.
 Note Eq.(\ref{pinningpot}) is similar to the one
 for SDW \cite{tmtsf1,tmtsf2} except for the $x$ integral coming from the $\bf k$ dependence of the 
 gap.
Then following FLR \cite{fl,lr}, in the strong pinning limit the threshold electric field at $T=0K$ is given by
\begin{equation}
E_T^S(0)=\frac{2k_F}{e}\frac{n_i}{n} N_0^2V_0V_y\frac{16}{\pi}0.5925
\Delta(0),
\end{equation}
and for general temperature it is obtained as
\begin{equation}
\frac{E_T^S(T)}{E_T^S(0)}=\frac{\rho_s}{\rho_s(T)}\frac{\Delta(T)}{\Delta(0)}
\frac{1}{0.5925} \int_0^1\tanh\frac{\beta\Delta(T)
x}{2}E(\sqrt{1-x^2})(K(x)-E(x))dx. \label{kuszobter}
\end{equation}
At low temperature $E_T^S$ increases linearly with $T$ since $\rho_s(T)$ is
linear in this range:
\begin{equation}
\frac{E_T^S(T)}{E_T^S(0)}=1+2\ln 2 \frac{T}{\Delta(0)},
\end{equation}
and the other quantities change like $T^3$.
At $T_c$, Eq. \ref{kuszobter} gives
\begin{equation}
\frac{E_T^S(T_c)}{E_T^S(0)}=\frac{\pi^3}{7\zeta(3)}\left(\frac{2\pi}{\sqrt
e\gamma}\right)^{-1}\frac{2\pi^2}{32\times0.5925}\approx 1.793,
\end{equation}
where $\gamma=1.781$. Close to the transition temperature $E_T$ increases
linearly:
\begin{equation}
\frac{E_T^S(T)}{E_T^S(0)}=\frac{E_T^S(T_c)}{E_T^S(0)}\left(1-0.42\left(1-\frac{T}{T_c}\right)\right).
\end{equation}
With its $T=0K$ slope, the normalized threshold field would reach 1.64 at
$T_c$, so it is almost linear in the strong pinning limit.

\begin{figure}
\hspace*{2cm}
\psfrag{x}[t][b][1][0]{$\frac{T}{T_c}$}
\psfrag{y}[b][t][1][0]{$\frac{E_T(T)}{E_T(0)}$}
{\includegraphics[width=13cm,height=8cm]{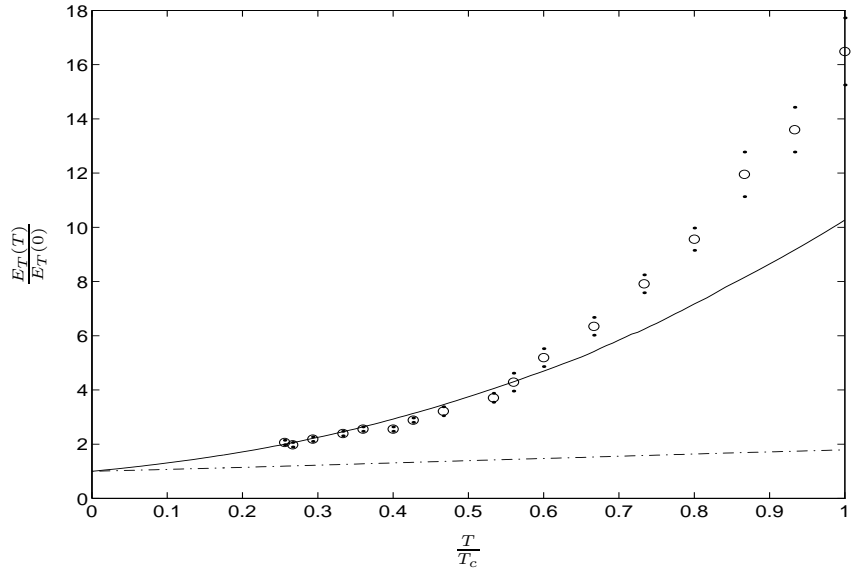}}
\\
\caption{The normalized threshold field plotted as a function of the reduced 
temperature in the strong pinning (solid line), weak-pinning (dashed-dotted 
line) limit. The circles are the measured values in $\alpha-(ET)_2KHg(SCN)_4$,
the points are the error bars.\label{fig:terek}}
\end{figure}

The strong pinning limit implies that the pinning potential is so strong
that one single impurity is adequate to pin the UCDW.
On the other hand, unless impurities are introduced by X-ray irradiation or
by some violent means, the weak-pinning limit appears to prevail \cite{tmtsf3}.
 Then
for the 3D weak-pinning limit we obtain \cite{fl,lr}
\begin{equation}
\frac{E_T^W(T)}{E_T^W(0)}=\left(\frac{E_T^S(T)}{E_T^S(0)}\right)^4.
\end{equation}
The threshold field is shown in Fig. \ref{fig:terek} together with the
data taken from Ref. \cite{ltp}. We see that the 3D weak-pinning limit is
qualitatively consistent with the experimental data. In other words, 
unconventional CDW appears to describe the LTP of $\alpha-(ET)_2KHg(SCN)_4$.
Also the present result applies also for unconventional SDW. On the other hand 
there 
is obvious discrepancy as $T$ approaches $T_c$.
In a forthcoming paper, we shall discuss the effect of imperfect nesting in  
order to improve the agreement between experiment and theory 
due to the fact that the $\alpha-(ET)_2$ salts'
Fermi surface contains two dimensional parts as well.

\section{Concluding remarks}

Within the theoretical framework developed in \cite{nagycikk} we study the
threshold electric field of unconventional CDW. The present result for the
3D weak-pinning limit appears to describe the data taken from the LTP of
$\alpha-(ET)_2KHg(SCN)_4$ satisfactorily. For this we need impurities with
anisotropic scattering amplitude \cite{haran,maki}. Together with the $H-T$ phase diagram
which is very parallel to the FFLO state in UCDW, the present result
indicates strongly that the LTP of $\alpha-(ET)_2MHg(SCN)_4$ with $M=K$,
$Tl$ and $RB$ is of unconventional CDW. In this respect a further study of
the threshold electric field in the presence of magnetic field will be of
great interest.

\begin{acknowledgments}
We thank Bojana Korin-Hamzi\'c for sending us the experimental data before
publication which provided us with initial stimulus to undertake this work. We thank also
Mark Kartsovnik for useful discussions. This work
was supported by the Hungarian National Research Fund under grant numbers
OTKA T032162 and T029877, and by the Ministry of Education under grant number
FKFP 0029/1999.
\end{acknowledgments}
\bibliography{et}

\end{document}